\documentstyle[aps,psfig,epsf]{revtex}
\author{
Mario Nicodemi 
}
\address{
\vspace{0.2cm}
Dipartimento di Fisica, Universit\'a di Napoli ``Federico II'',
INFM and INFN Sezione di Napoli \\
Mostra d'Oltremare, Pad. 19, 80125 Napoli, Italy\\
P.M.M.H. Ecole de Physique et Chimie Industrielles,
10, rue Vauquelin, 75231 Paris CEDEX 05 France \\
}

\title{ Force correlations and arches formation in granular assemblies }

\date{\today}

\newcommand{\lan}{\langle}
\newcommand{\ran}{\rangle}
 
\begin{document}

\maketitle
\bigskip
\begin{abstract}
In the context of a simple microscopic schematic scalar model we study
the effects of spatial correlations in force transmission
in granular assemblies. We show that the parameters of the
normalized weights distribution function, $P(v)\sim v^{\alpha}\exp(-v/\phi)$,
strongly depend on the spatial extensions, $\xi_V$, of such correlations.
In particular we evaluate the functions $\phi(\xi_V)$ and $\alpha(\xi_V)$.
If $\xi_V\rightarrow\infty$ then $\phi\sim \xi_V$ and weights are power
law distributed.
\end{abstract}
\bigskip

Photoelastic visualizations of stress propagation in granular media 
have provided evidence of strong inhomogeneities in the distribution 
of contact forces with the formation of ``force chains" which typically 
extend on space scales much larger than the dimensions of 
single grains \cite{Dantu-Travers-Drescher,JNScience95,Behringer96}. 
These inhomogeneities results from local fluctuations and are accompanied
by broad stress distributions observed in bead packs 
\cite{JNScience95,Behringer96,JNBHM} which are responsible for 
many unusual properties of granular media 
%\cite{JNBHM,Liu-Smid,Savage}. 
\cite{JNBHM}. 

Several theoretical approaches have been proposed to
describe the unusual properties of stress patterns in such systems
\cite{JNBHM,EdwardsMounfield,Mehta92,BouchaudCatesClaudin} and 
recently a schematic scalar model, which takes into account just the vertical
component of the stress tensor, has been introduced to describe the 
statistics of stress distribution \cite{JNScience95,Coppersmith}.
The mean field study of this model has shown rich results in good 
correspondence with experimental observations. However recent experiments
\cite{Behringer96} show the necessity to go beyond the mean field 
approximation and to take into account spatial correlations present 
in force transmission from grain to grain, correlations 
which actually lead to chains formation.  

In the present paper, we study the effects of such correlations in the 
framework of a schematic scalar model, as in
ref.~\cite{JNScience95,Coppersmith}, in which grains, disposed on a lattice,
discharge their weights in random fractions, due to the disorder
present in a typical granular assembly, on their bottom neighbours. 
The disorder of the environment, however, does not imply that 
force transmission between grains is uncorrelated 
\cite{EdwardsMounfield,ClaudinBouchaud,NCH}. 
In the complex dynamical process which leads an assembly to
a static configuration, a certain fraction of sites carrying a strong
shear (typically involved in the formation of ``force chains''), 
slips away from some of their own neighbours, losing contact with them. 
These processes may induce strong spatial correlations in force transmission, 
which we try to take into account here. 

In our model the weight,
$w(i,h)$, supported by a grain at height $h$ and column $i$ of a square
lattice (see Fig.~\ref{chains}) is 
(using the notation of ref.~\cite{Coppersmith,ClaudinBouchaud}):
\begin{eqnarray}
w(i,h)&=& q_+(i-1,h-1)~w(i-1,h-1) \nonumber \\
&&+q_-(i+1,h-1)~w(i+1,h-1)+1
\label{trans_eq}
\end{eqnarray}
where $q_+$ (resp. $q_-$) is the fraction of the weight that site
$(i-1,h-1)$ (resp. $(i+1,h-1)$) discharges on site $(i,h)$, and
we have supposed that the masses of single grains are equal to unity. 
Due to the conservation of the mass, 
we have the constraint $q_+(i,h)+q_-(i,h)=1$. 
The values of $q_+(i,h)$ are generally uniformly distributed in the
interval $[0,1]$. However, we suppose that a fraction $1-\delta$ 
($\delta\in[0,1]$) of the total number of grains
(a fraction of grains randomly displaced on the lattice) 
is subjected to the slip condition proposed in ref.~\cite{ClaudinBouchaud}:
\begin{equation}
q_+(i,h)=\theta(x)
\label{teta}
\end{equation}
Here $\theta$ is the Heaviside step function and $x$ is \cite{nota_soglia}: 
\begin{eqnarray}
x &=&\left[ q_+(i-1,h-1)~w(i-1,h-1)\right. \nonumber \\
&&\left.-q_-(i+1,h-1)~w(i+1,h-1)\right]/w(i,h)
\label{slip_cond}
\end{eqnarray}
Eq.~(\ref{teta}) schematically expresses the slip condition 
in which, due to friction, a grain, strongly ``pressed" from its top right 
(resp. left) neighbour, may discharge its weight mainly on its own 
right (resp. left) lower neighbour \cite{robustness}. 
This mechanism originates force chains in our model. 

Our tuning parameter is the fraction of grains not subject to the slip 
condition, $\delta$, a parameter which, in real samples, is related to 
friction coefficients, elastic constants, grains shapes. The 
model interpolates the two extreme situations corresponding to a 
complete uncorrelated weight transmission, studied in 
ref.~\cite{Coppersmith}, which is obtained when $\delta=1$,  
and the strongly correlated case studied in  ref.~\cite{ClaudinBouchaud}, 
which is recovered in the limit $\delta=0$. 

This two dimensional model is simple enough to visualize  
mechanisms underlying force transmission. Actually, 
the sites which undergo the above ``slip condition''
are typical points enhancing correlations in force transmission and so 
belonging to ``force'' chains (see Fig.~\ref{chains}). 
In our case the ``slipping'' grains are randomly distributed on the lattice; 
thus the probability to have a segment of such grains of total length $d$, 
i.e., to have a chain of length $d$,
is approximately ${\cal P}(d)\sim (1-\delta)^d$. 
Consequently the average chain length, $\xi_V$, which is 
experimentally easier to measure, is related 
to the fraction of such sites as: $\xi_V\simeq (1-\delta)/\delta$. 
The kind of correlation we have introduced above essentially
generates ``chains'' which sustain the weight of the lattice columns
which intersect them from above, as ``arches" 
\cite{EdwardsMounfield,BouchaudCatesClaudin}. Thus, in the deep bulk
of the system at depth $L$, the weight, $W$, that a point
at the base of a chain of length $d$ experiences, is approximately  
$W\sim Ld$. 
A quantity of theoretical as well as experimental and practical importance 
is the distribution $P(v)$ of normalized weights, $v=W/L$, felt by grains
at a given depth $L$. With the above arguments we can derive an
approximate expression to relate $P(v)$ 
to the fraction $\delta$, i.e., to the average chain length $\xi_V$:
\begin{equation}
P(v)\equiv P(W(d)/L)={\cal P}(d)\sim \exp(-v/\phi)
\label{dist_theor}
\end{equation}
where $\phi\sim 1/\log[1/(1-\delta)]$ ($\phi\sim \xi_V$, if 
$\xi_V\rightarrow\infty$). 
This result implies important practical consequences, because it predicts 
that we have to expect to measure weight fluctuations of the order of 
$\xi_V$. Thus if $\xi_V$, which is a characteristic length of the specific 
system we consider, 
is very high, huge stress fluctuations arise in the sample. 
The above picture compares well with the numerical calculations
we present below, where its consequences are discussed in more details. 

Our numerical analysis of the present model concerns a square lattice 
of length $L=10000$ and depth $M=1000$, averaged over $10$ different 
realizations. In this lattice we adopted periodic boundary conditions 
along the horizontal axis. 

To study the effects of spatial correlations introduced by the slip 
condition of eq.~(\ref{teta}), we evaluate the vertical 
space correlation function of forces, $C_V(r)$, along a main axis of our
square lattice:
\begin{equation}
C_V(r)= \frac{ \lan w(i,M) w(i-r,M-r) \ran - w_m(0)w_m(r)}
{\lan w(i,M)^2 \ran - w_m(0)^2} 
\label{cor_V}
\end{equation}
with $w_m(r)=\lan w(i,M-r)\ran$. 
The behaviour of $C_V(r)$, in the present model, is depicted in
Fig.~\ref{corr} as a function of the distance between grains $r$ 
(expressed in lattice units). 
After a first jump from $C_V(0)=1$ to $C_V(1)\sim 0.5$, 
$C_V(r)$ smoothly decreases approaching zero. 
The first part of the decay is exponential in $r$, and 
may be fitted with the following function:
\begin{equation}
C_V(r)=K_V \exp(-r/\xi_V)
\label{exp_cor_V}
\end{equation}
The characteristic length $\xi_V$ 
of eq.~(\ref{exp_cor_V}), is related to the 
measure of the extensions of forces inhomogeneities along 
the vertical direction in our system. It is the typical vertical 
extension of forces correlation, or, more crudely, the
average vertical distance between crossing points of ``stress paths'' 
observed in photoelastic measurements (see Fig.~\ref{chains}). 
In Fig.~\ref{cor_param}, $\xi_V$ is plotted as a
function of $\delta$ \cite{nota_Cv}. 
In agreement with the above theoretical arguments, 
the approximate behaviour $\xi_V\sim \delta^{-1.0}$ is
found, showing that $\xi_V$ increases orders 
of magnitude when the fraction of sites, $1-\delta$,  
undergoing the slip condition approaches $1$. 
In the limit
$\xi_V\rightarrow\infty$, force chains extends over the whole 
system and this fact strongly affects weights distribution. 

The horizontal space correlation function, $C_H(r)$, is defined as
the average over the lower $10\%$ of the system (to have
more precise data) of $c(r,h)$: 
\begin{equation}
c(r,h)= \frac{\lan w(i,h) w(i+r,h) \ran - \lan w(i,h) \ran^2}
{\lan w(i,h)^2 \ran - \lan w(i,h) \ran^2} ~ ~ .
\end{equation}
The function $C_H(r)$, depicted in
Fig.~\ref{corr}, becomes negative as soon as 
$r\geq 1$, signaling that horizontally aligned grains are always slightly 
anti-correlated. When $r\rightarrow\infty$, $C_H(r)$ 
asymptotically approaches zero from below. As above, the first part 
of the decay is almost exponential:
\begin{equation}
C_H(r)=-K_H \exp(-r/\xi_H)
\label{exp_cor_H}
\end{equation}
In this case, the length $\xi_H$ corresponds to the typical 
horizontal spacing of chains. The function $\xi_H(\xi_V)$ is reported in
Fig.~\ref{cor_param}. $\xi_H$ diverges  
when $\delta\rightarrow 0$, and we approximately find
$\xi_H\sim\xi_V^a$ with $a\sim 1/2$.  

We now try to relate the above observations to 
the distribution, $P(v)$, of the 
weights, $v=w/w_m$, normalized by the mean $w_m=h$, 
supported by grains at a given depth $h$. 
The function $P(v)$, evaluated at the bottom layers in our model, is 
plotted in Fig.~\ref{dist}. In agreement with the above theoretical
considerations, this quantity becomes independent of 
$h$ if measured at sufficient depth in the system.
In ref.~\cite{JNScience95,Coppersmith} has been proposed that $P(v)$
is characterised, in a very broad class of models, 
by the following behaviour, which is approximately recovered in the 
present model (we find deviations at very small $v$): 
\begin{equation}
P(v)=A v^{\alpha} \exp(- v/\phi)
\label{vexpv}
\end{equation}
An interesting result from our numerical calculation, consistent with
the theoretical considerations presented above, is that the asymptotic 
exponential behaviour of $P(v)$ is recovered for all 
values of $\delta > 0$. The correlations present in force transmission
between grains, except exceptional cases, do not alter this property. 
The validity of this important observation may be broader, being  
consistent with results obtained in different contexts 
as the carbon paper experiment with compressed stationary bead packs 
of ref.~\cite{JNScience95} or experiments 
on continuously sheared granular materials of ref.~\cite{Behringer96}, 
or in numerical simulations of scalar or vectorial force models 
\cite{Coppersmith,NCH,Roux,Bagi}. 

However, the quantities $\alpha$ and $\phi$, which in mean field theory 
seem to be exclusively related to the coordination number of the 
lattice of grain packing \cite{Coppersmith,NCH}, strongly depend 
on the degree of correlation present in forces transmission, i.e., 
in our model, on the fraction $\delta$. This fact has important
practical consequences and may explain contrasting results found in 
different experiments. 
The dependence of the distribution $P(v)$ on spatial correlations 
has been recently experimentally outlined in ref.~\cite{Behringer96}. 

In the present model, the exponent $\alpha$ decreases in presence of 
spatial correlations in force transmission between grains. It passes from the 
value $\alpha=1$ at $\delta=1$ (as predicted by mean field theory) to 
the value $\alpha\simeq -1.1$ at $\delta=0$ (in agreement with the simulations
of ref.~\cite{ClaudinBouchaud}). At a fraction of 
``normal'' sites $\delta\sim 0.7$, a value higher than the percolation
threshold on the square lattice, $\alpha$ crosses the zero. 
The exponent $\alpha$ is depicted in Fig.~\ref{dis_param} as a
function of the vertical correlation length $\xi_V$. 
% The crude fit shown in the figure is: 
% $\alpha=\alpha_0/(\xi_V/\xi_0-1)^{b}-1.1$, with 
% $\alpha_0=1.3$, $\xi_0=1.2$, and $b=0.5$. 
Figure~\ref{dis_param} 
shows that in the region of not too large values of $\xi_V$, 
small changes in $\xi_V$ may induce large variations of $\alpha$. 
Such a strong dependence seems to be found also in the experimental 
observations of ref.~\cite{Behringer96}, where was shown 
the sensitivity of the measure of $\alpha$ with respect to the
relative sizes of the grains and of 
the measuring device. In particular, from the above general
discussion and the results of ref.~\cite{Behringer96}, 
we expect that the results of mean field theory have to be 
experimentally recovered essentially when the measures of forces are taken
averaging over regions which are larger than the length
$\xi_V$. 

For what concerns the parameter $\phi$, 
Fig.~\ref{dis_param} shows that it diverges approximately as a power law
of $\xi_V$. The superimposed fit, consistent with the previous theoretical 
arguments, is $\phi(\xi_V)\sim \xi_V^{1.0}$. 
Thus $\phi$ grows with $\xi_V$ and in the limit of huge spatial correlation, 
i.e., $\xi_V\rightarrow \infty$ (or $\delta\rightarrow 0$), 
the exponential asymptotic decay of eq.~(\ref{vexpv}) is lost and 
just the power law behaviour survives. 
This fact has significant practical importance because it 
implies that giant stress fluctuations 
can be observed \cite{ClaudinBouchaud}.

In conclusion, we have studied a simple microscopic model in the framework of 
scalar force approximation \cite{Coppersmith,ClaudinBouchaud}, which 
schematically takes into account the effects of correlations present 
in forces transmission in granular assemblies. 
Its simplicity allows to clarify the effects induced by 
the extension of ``forces chains" in the system. 
The presence of such a non trivial characteristic length scale, 
which is experimentally measured for 
instance by photoelastic visualization, can thus be quantitatively related 
to other measurable quantities as the forces distributions, $P(v)$. 
We have discussed some interesting correspondences with known results 
from recent experiments \cite{Behringer96}. However further experimental 
and theoretical investigation about these important effects is still 
missing. 

Interestingly the present model is related (see ref.~\cite{Coppersmith}) 
to works devoted to describe other physical phenomena as directed Abelian 
sandpiles, aggregation-dissociation reactions, interface dynamics, 
or river networks \cite{Takayasu-Dhar}. 

{\em Acknowledgements} It's a pleasure to thank Prof. R. P. Behringer
for very fruitful discussions and his critical reading of the manuscript.

%\newpage

\begin{figure}[h]
%\centerline{
%  \psfig{figure=fig_chains_08.ps,width=4cm,angle=0}
%  $~~~~~~$
%  \psfig{figure=fig_chains_01.ps,width=4cm,angle=0}
%}
%\vspace{0.3cm}
\caption{
Two pictures of the conformation of stress paths in our model, corresponding 
to two values of the correlation length, $\xi_V$ ($\xi_V\sim 2$ in the left 
picture, and $\xi_V\sim 20$ on the right). They are obtained enlightening 
the grains which carry a weight above a given threshold. 
} 
\label{chains}
\end{figure}

\begin{figure}[h]
%\centerline{\psfig{figure=fig_cor.ps,width=6cm,angle=-90}}
%\vspace{0.5cm}
\caption{
The vertical, $C_V(r)$, and the horizontal, $C_H(r)$, force correlation
functions as a function of the grain distance $r$ (in unit
of the lattice spacing), for several values of the fraction of grains
not undergoing the ``slip condition'' of eq.~(\ref{teta}), 
$\delta=0.1,0.3,0.6,1.0$ (resp. squares, triangles, diamonds,
circles). The superimposed curves are the exponential fits described
in the texts.   
} 
\label{corr}
\end{figure}

\begin{figure}[h]
%\centerline{\psfig{figure=fig_dati_cor.ps,width=6cm,angle=-90}}
%\vspace{0.5cm}
\caption{
{\em Left:} the characteristic length $\xi_V$ of the vertical force 
correlation function as a function of $\delta$. 
It diverges when $\delta\rightarrow 0$
approximately as $\xi_V\sim \delta^{-1.0}$. In such a limit
force correlations extend over the whole system, and this fact strongly
affects forces distribution. 
{\em Right:} the characteristic length $\xi_H$ of the horizontal 
force correlation function as a function of $\xi_V$ ($\xi_H\sim \xi_V^{1/2}$). 
} 
\label{cor_param}
\end{figure}

\begin{figure}[h]
%\centerline{\psfig{figure=fig_dis.ps,width=6cm,angle=-90}}
%\vspace{0.5cm}
\caption{
The distribution $P(v)$ of 
forces $v=w/w_m$ normalized by the mean force $w_m$ at the bottom of
our system, depicted for three values of the fraction of grains
not undergoing the ``slip condition'' of eq.~(\ref{teta}), 
$\delta=0.1,0.6,1.0$ (resp. squares, diamonds, circles). 
The superimposed fits are from eq.~(\ref{vexpv}). 
} 
\label{dist}
\end{figure}

\begin{figure}[h]
%\centerline{\psfig{figure=fig_dati_dis_in_funz_cor.ps,width=6cm,angle=-90}}
%\vspace{0.5cm}
\caption{
The parameters of eq.~(\ref{vexpv}), $\alpha$ and $\phi$, 
as a function of the correlation length $\xi_V$. 
{\em Right:} the exponent $\alpha$ passes from the 
value predicted by mean field theory, $\alpha=1$, at small $\xi_V$ 
(i.e., $\delta=1$) to $\alpha\simeq -1.1$ when $\xi_V\rightarrow\infty$
(i.e., at $\delta=0$). The sensitivity of $\alpha$ to changes of
$\xi_V$ remembers the observations from experiments by Miller et al. [3]. 
{\em Left:} the parameter $\phi$ diverges as a power law with $\xi_V$
(approx. $\phi \sim \xi_V$), showing that if $\xi_V\rightarrow \infty$ 
the exponential asymptotic decay of force distribution $P(v)$ is lost, and 
huge stress fluctuations are possible. 
} 
\label{dis_param}
\end{figure}


\bigskip

\begin{thebibliography}{40}

\bibitem{Dantu-Travers-Drescher} P. Dantu, in {\em Proc. of the 4th 
Int. Conf. on Soil Mech. and Found. Eng.}
(Butterworths, London, 1957). 
A. Drescher, G. De Josselin De Jong, J. Mech. Phys. Solids {\bf 20},
337 (1972). T. Travers, D. Bideau, A. Gervois, J. C. Messager,
J. Phys. A {\bf 19}, L1033 (1986).

\bibitem{JNScience95} C.-h. Liu, S.R. Nagel, D.A. Schecter,
  S.N. Coppersmith, S. Majumdar, O. Narayan, T.A. Witten, Science {\bf 269},
 513 (1995).

\bibitem{Behringer96} B. Miller, C. O'Hern and R. P. Behringer,
  Phys. Rev. Lett. {\bf 77}, 3110 (1996).

\bibitem{JNBHM} H.M. Jaeger and S.R. Nagel, Science {\bf 255},
  1523 (1992);
  H.M. Jaeger, S.R. Nagel and R.P. Behringer,
  Rev. Mod. Phys. {\bf 68}, 1259 (1996).

%\bibitem{Liu-Smid} C.-h. Liu and S. R. Nagel, 
%  Phys. Rev. Lett. {\bf 68}, 2301 (1992).
%J. Smid and J. Novosad, in Proc. of 
%{\em 1981 Powtech. Conf., Ind. Chem. Eng. Symp.} {\bf 63}, D3V 1 (1981). 

%\bibitem{Savage} 
%S. B. Savage, Adv. Appl. Mech. {\bf 24}, 289 (1984).
%C. S. Campbell, Ann. Rev. Fluid Mech. {\bf 22}, 57 (1990). 

\bibitem{EdwardsMounfield} 
S.F. Edwards and R. B. Oakeshott, Physica D {\bf 38}, 88 (1989). 
S.F. Edwards and C.C. Mounfield, Physica A 
{\bf 226}, 1 (1996); ibid. {\bf 226}, 12 (1996); ibid. {\bf 226}, 25
(1996).

\bibitem{Mehta92} A. Mehta, Physica A {\bf 186}, 121 (1992).

\bibitem{BouchaudCatesClaudin} 
J.-P. Bouchaud, M.E. Cates and P. Claudin, 
  J. Physique {\bf I5}, 639 (1995).
J. P. Wittmer, P. Claudin, M.E. Cates, J.-P. Bouchaud, 
Nature {\bf 382}, 336 (1996). 

\bibitem{Coppersmith} 
S.N. Coppersmith, C.-h. Liu, S. Majumdar, O. Narayan, T.A. Witten,
  Phys. Rev. E {\bf 53}, 4676 (1996).

\bibitem{ClaudinBouchaud} P. Claudin and J.-P. Bouchaud, 
Phys. Rev. Lett. {\bf 78}, 231 (1997).

\bibitem{nota_soglia} 
In ref.~\cite{ClaudinBouchaud} also a phenomenological threshold, 
${\cal R}_c\in[0,1]$, was introduced. However, except when 
${\cal R}_c$ is close to one, it doesn't play a major role 
so we fix ${\cal R}_c=0$. 

\bibitem{robustness} The main results we present here 
are well robust to some changes of the model. Without finding 
interesting 
alterations, we explored several versions of 
conditions (\ref{teta}) and (\ref{slip_cond}). We
also applied them to the model of ref.~\cite{NCH}, where the 
disorder of the $q_{\pm}$ is substituted by 
disorder in the position of holes in grains packs. 

\bibitem{NCH} M. Nicodemi, A. Coniglio, H.J. Herrmann, 
J. Phys. A {\bf 30}, L379 (1997); Physica A {\bf 240}, 405 (1997). 

%\bibitem{nota_an} This can be analytically proved with the same techniques 
%introduced in ref.~\cite{Coppersmith}. 

\bibitem{Roux} F. Radjai, M. Jean, J.-J. Moreau and S. Roux, Phys.
Rev. Lett. {\bf 77}, 274 (1996).

\bibitem{nota_Cv} The function $C_V(r)$ when
$r\geq \xi_V$ may become slightly negative before approaching zero
in the limit $r\rightarrow\infty$.

\bibitem{Bagi} K. Bagi, in {Powders and Grains 93}, edited by C. Thorton 
(Balkema, Rotterdam, 1993). 

\bibitem{Takayasu-Dhar} H. Takayasu, I. Nishikawa and H. Tasaki, 
Phys. Rev. A {\bf 37}, 3110 (1988). D. Dhar and R. Ramaswamy, Phys. Rev. Lett. 
{\bf 63}, 1659 (1989). 

\end{thebibliography}
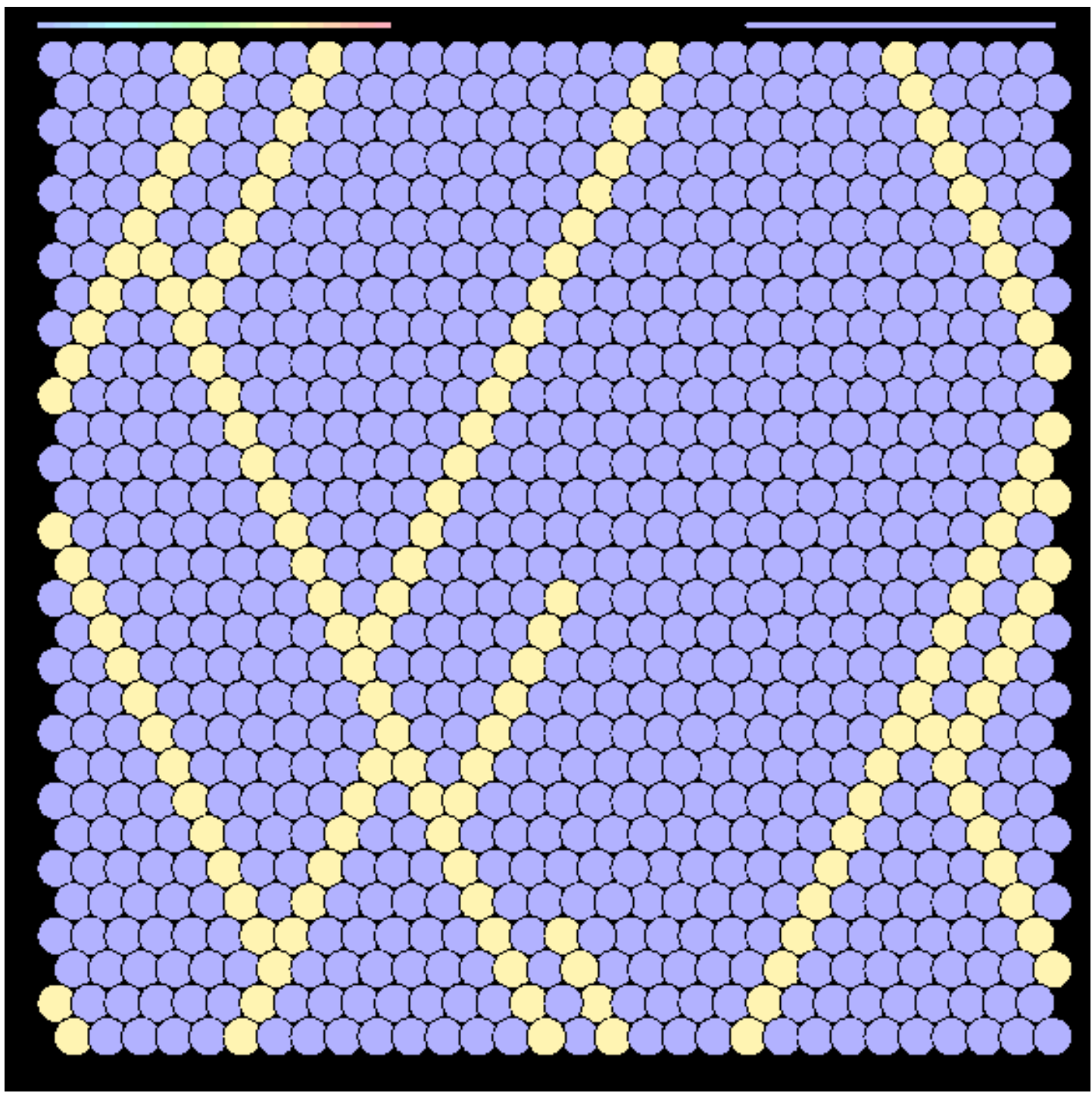
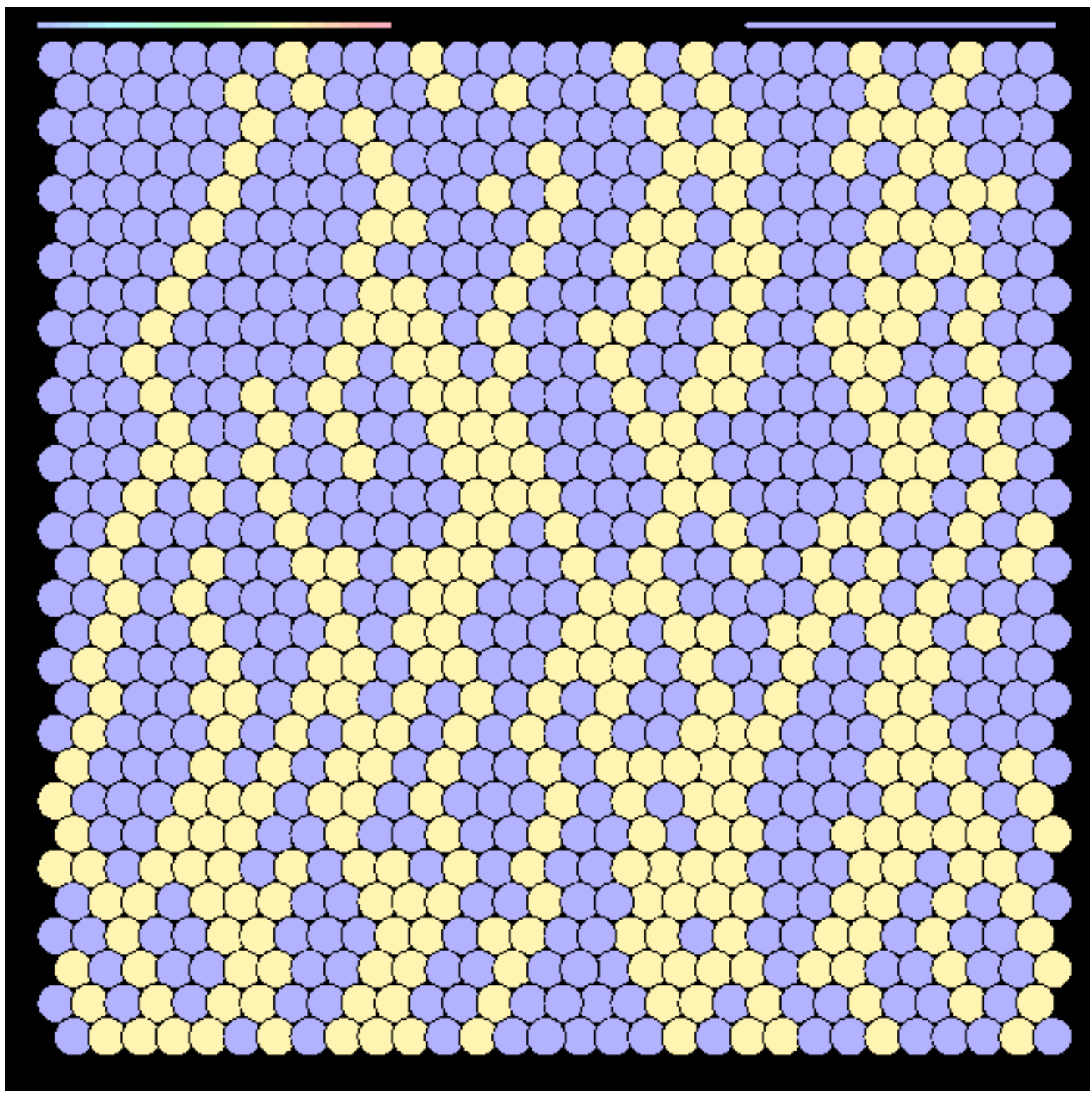
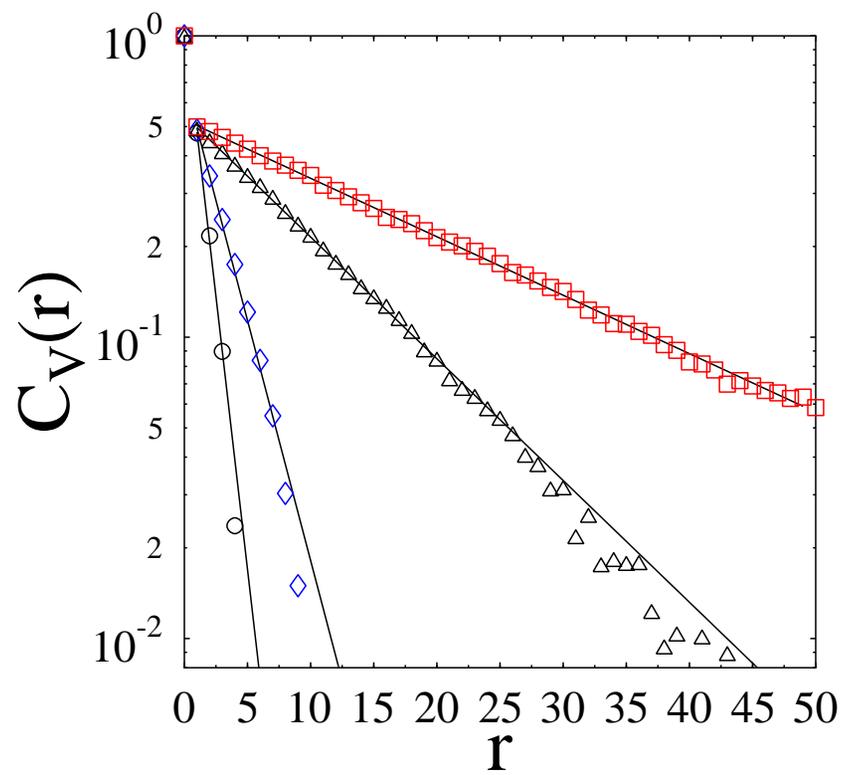
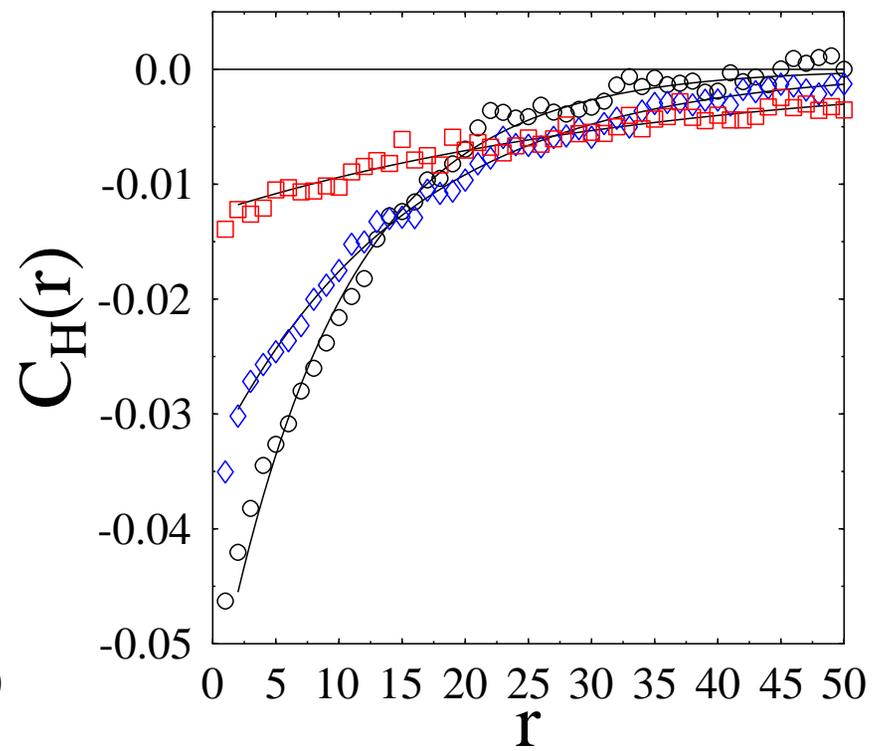
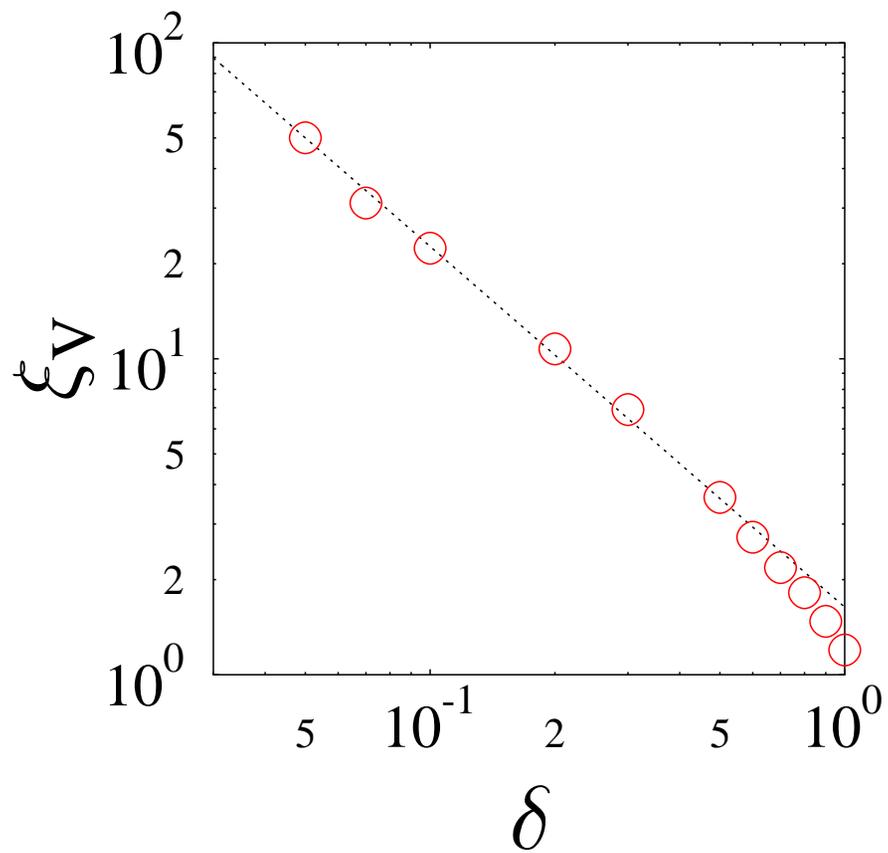
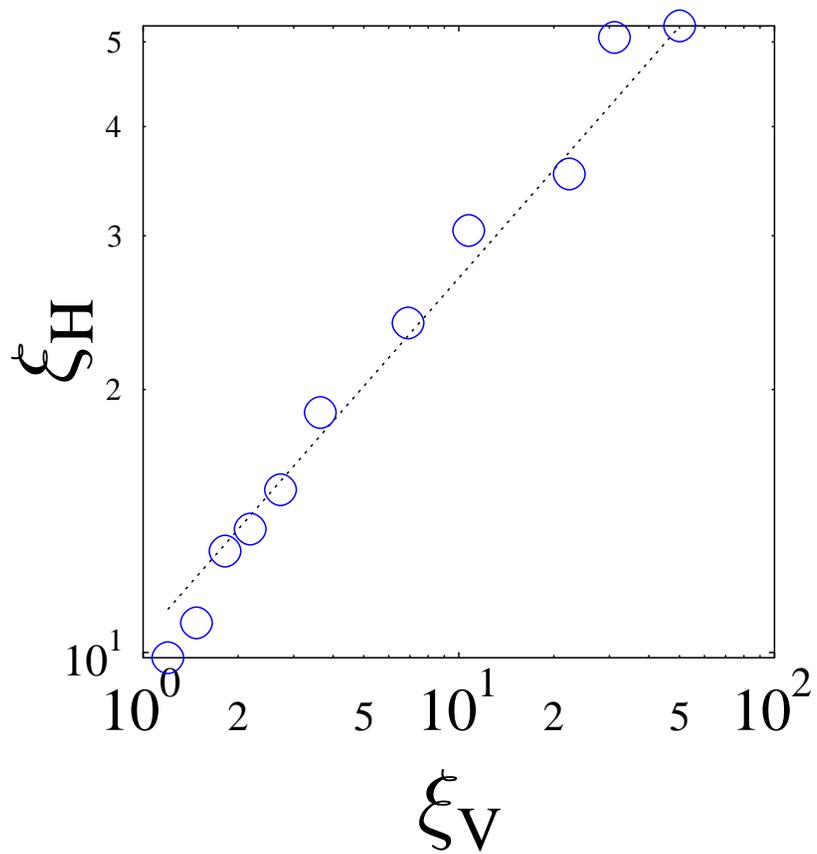
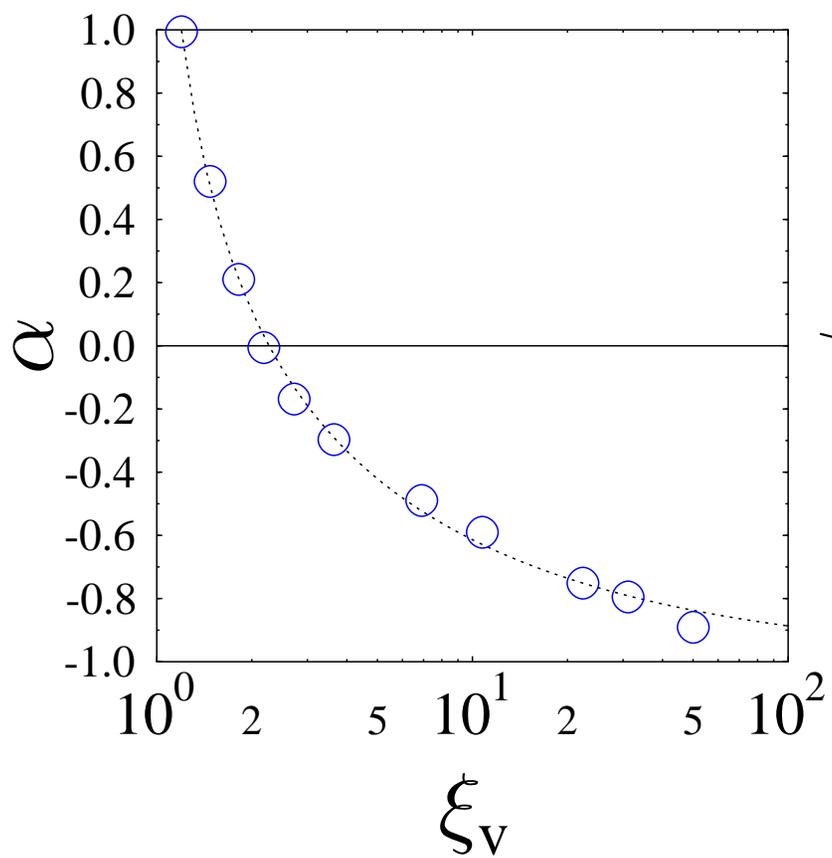
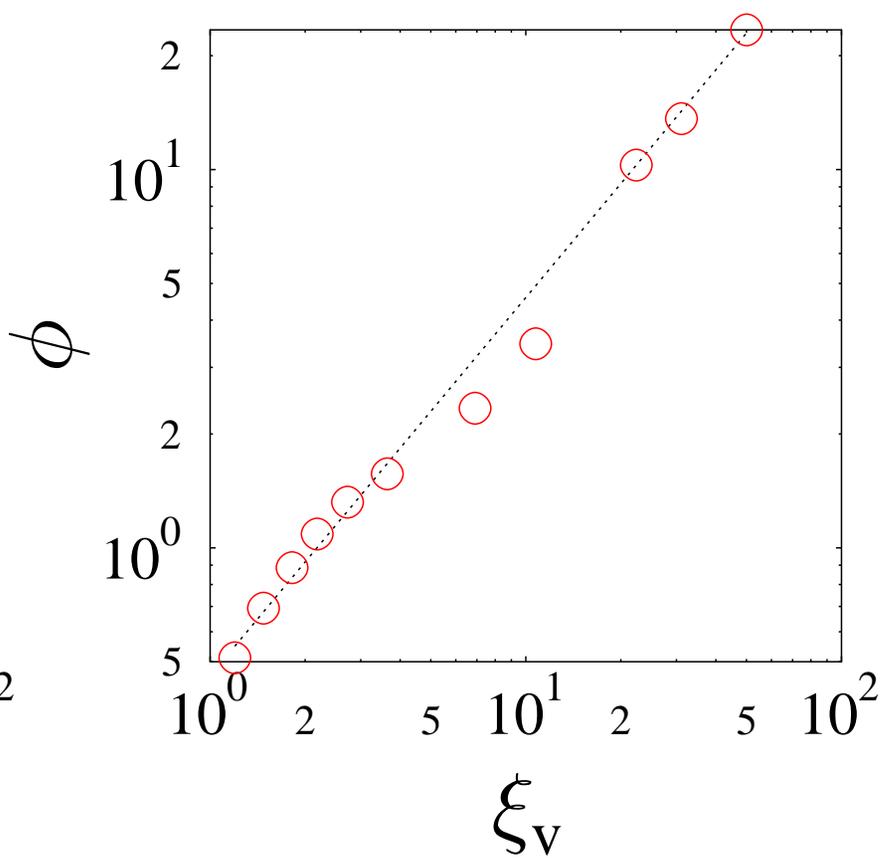
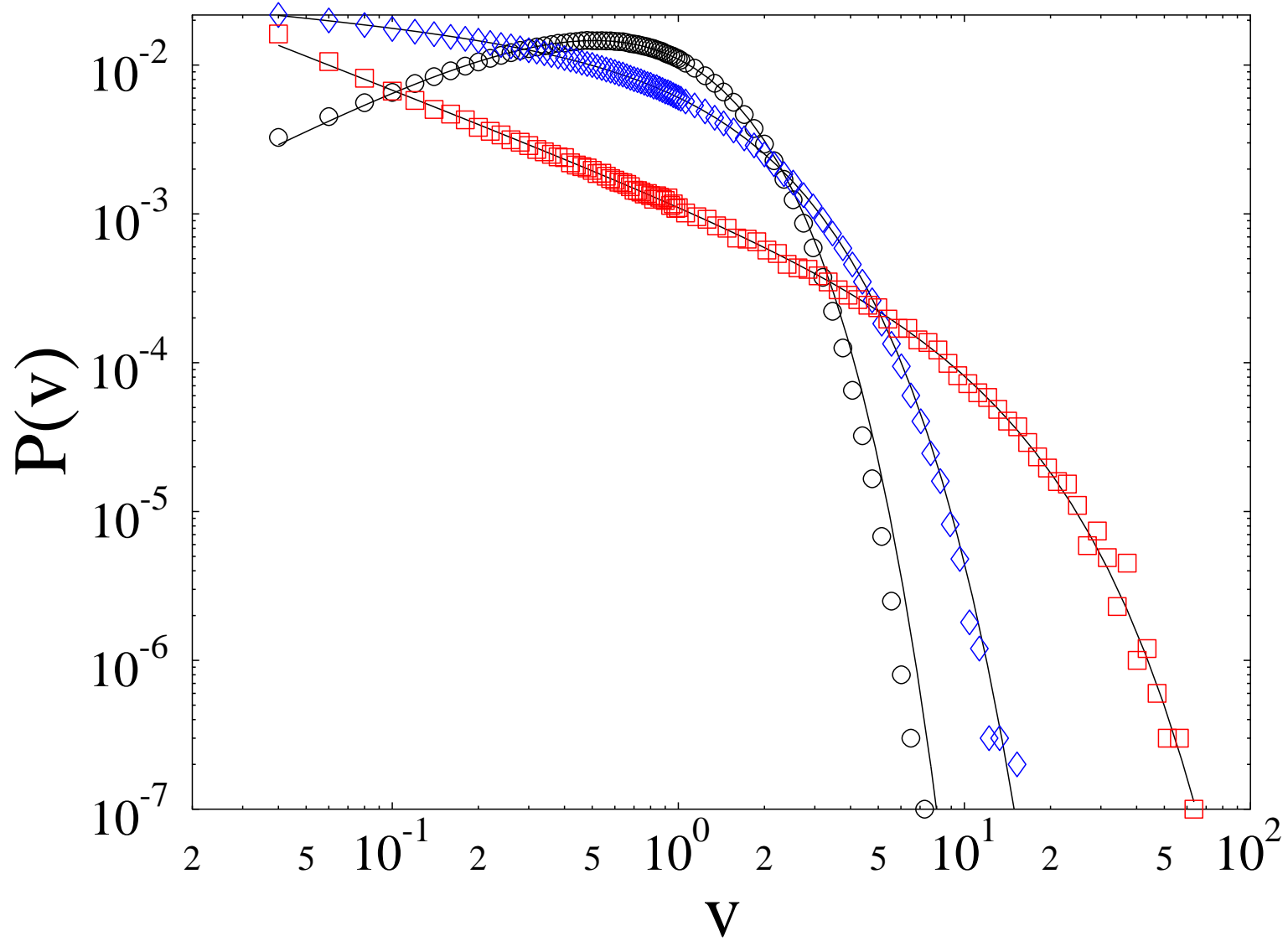
\end{document}